# All-Materials-Inclusive Flash Spark Plasma Sintering

Charles Manière, Geuntak Lee, Eugene A. Olevsky*

Powder Technology Laboratory, San Diego State University, San Diego, USA

**Abstract**

A new flash (ultra-rapid) spark plasma sintering method applicable to various materials systems, regardless of their electrical resistivity, is developed. A number of powders ranging from metals to electrically insulative ceramics have been successfully densified resulting in homogeneous microstructures within sintering times of 8-35 s. A finite element simulation reveals that the developed method, providing an extraordinary fast and homogeneous heating concentrated in the sample's volume and punches, is applicable to all the different samples tested. The utilized uniquely controllable flash phenomenon is enabled by the combination of the electric current concentration around the sample and the confinement of the heat generated in this area by the lateral thermal contact resistance. The presented new method allows: extending flash sintering to nearly all materials, controlling sample shape by an added graphite die, and an energy efficient mass production of small and intermediate size objects. This approach represents also a potential venue for future investigations of flash sintering of complex shapes.

---

* Corresponding author: **EO**: Powder Technology Laboratory, San Diego State University, 5500 Campanile Drive, San Diego, CA 92182-1323,
Ph.: (619)-594-2420; Fax: (619)-594-3599, *E-mail address*: eolevsky@mail.sdsu.edu

# Introduction

The quest for a stable sintering process able to gather high material performance, high energy efficiency and short operating time lends impetus to the discovery of new unconventional sintering techniques. The pressure assisted sintering techniques such as hot pressing or hot isostatic pressing enable 200-400 K lower processing temperatures than in conventional sintering and reductions of sintering time by few hours, which results in finer microstructures [1–3]. Significant improvements of sintering kinetics are also observed when the process is assisted by electric current [4] or electromagnetic field [5–8]. These powder consolidation technologies are predominantly represented in the modern studies by the electric current activated/assisted sintering (ECAS) techniques [9] which encompass resistive sintering (RS), where the sample is Joule heated by an externally applied voltage or AC/DC/pulsed electric current, and electrical discharge sintering (EDS), where high voltage electric discharge pulses are generated by a capacitor bank [10]. The spark plasma sintering (SPS) process is an ECAS technique that combines the benefits of highly pulsed electric current (1000-8000 A), high pressure (up to 100 MPa), high temperatures (up to 2400 °C), and high heating rates (up to 1,000 K/min) [3,11–13]. With these characteristics, a wide range of materials can be fully compacted, from the polymers to ultra-high temperature ceramics [14], such as TaC [15], $ZrB_2$ [16], etc. With more than 6,000 archival publications (as indicated by Web of Science®), the SPS process is now a well-established and stable technology allowing the fabrication of small and large size samples for sintering times between 5 and 60 min depending on the material and the targeted application [17]. Despite its advantages, the SPS technique is low productive because of its commonly utilized batch processing schematics. To improve its productivity two main approaches can be employed. One is based on the simultaneous sintering of multiple parts specimens [3]. The other solution, proposed in the present study, is the reduction of the sintering time to few seconds, combining SPS and flash sintering (FS) approaches.

Flash sintering is an energy efficient sintering process involving very short sintering times between 1 and 60 s [18]. The first flash sintering phenomenon has been discovered during the field assisted sintering studies of yttria stabilized zirconia [19]. The heating of zirconia under applied electric potential indicates the existence of the two main process stages. During the first incubation stage (which follows the preliminary stage of the external heating of the powder specimen by a conventional sintering furnace), the specimen is self-heated (in addition to the external heating) by an increasing amount of electric current passing through it. At an onset temperature related to the material resistivity behavior with temperature, the specimen becomes sufficiently conductive to allow a high electric current passage, and during this second stage a significant acceleration of the specimen's heating is accompanied by a very fast sintering in few seconds. This thermal runaway profile [20–22] enables heating rates of the order of 500-10,000 K/min and sintering times of few seconds [23,24]. This fast-resistive sintering method is easily applicable to many materials exhibiting a negative temperature coefficient (NTC) behavior of their electric resistivity like zirconia [25], silicon carbide [26], most of the thermistor materials like oxide spinel [27], perovskites [28], alumina (with MgO doping [29] or special electric field conditions [30]) and other materials [31,32].

The flash sintering may involve the following mechanisms. Joule heating [24,33] generates fast sample heating rates which may result in a thermal runaway [34] leading to accelerated sintering [35] as in experiments reported in Ref. [36]. The grain boundary melting or softening have been reported for microwave flash sintering [37] or for a traditional flash sintering process [38,39]. The nucleation of defects [40–42], local overheating of the grain boundaries [19] or the acceleration of sintering resulting from the dielectric breakdown when high electric fields are employed [43] are also possible mechanisms of flash sintering.

Similar FS approaches based on the microwave sintering, EDS, and SPS have been developed. The microwave sintering naturally demonstrates a tendency for high thermal [44] and sintering [37] runaway of ceramics that also stems from the NTC behavior of the specimens and/or tooling materials [45]. The EDS techniques use discharge times of the order of hundreds of microseconds and thus can be categorized as ultrafast sintering methods which are predominantly applied to conductive powders [10]. A similar ultrafast and contactless sintering method with discharge time between 2-10 s using plasma electrode [46] resulted in a nearly full densification of carbides (SiC, $B_4C$). A promising approach utilizing SPS in the sinter-forging configuration with graphite felt preheating element to initiate the flash phenomenon allowed the full densification of zirconia [47] and of high temperature materials such as SiC [48] or $ZrB_2$ [49] in about 10-20 s. A similar flash spark plasma sintering process can be achieved starting directly from a cold pressed powder with the noticeable improvement of the sample homogeneity [50]. An even faster approach using the sinter-forging SPS configuration [51] and SPS sacrificial tooling components succeeded to compact a powdered SiC sample in about 1 s.

Despite a strong potential, numerous technological issues of FS remain unsolved. The intrinsic instabilities responsible for the thermal runaway also generate local hot spots phenomena resulting in non-homogeneous microstructure developments [44,52,53]. FS employs dog bone shape specimens with platinum electrodes which reduces the commercial attractiveness of the method. Another issue is the increase of the overall operating time due to the preheating needed to activate the electric conductivity which is too low at room temperature for most of the NTC systems. But the greatest limitation of all the FS approaches is the lack of the applicability of FS to a wide range of materials. Contrary to NTC materials, metals and alloys have a very high conductivity that decreases with temperature preventing the appearance of a natural "intrinsic" thermal runaway under a constant applied voltage. In contrast, a high electric insulator, such as high purity alumina or boron nitride, does not allow the electric current to activate the Joule heating phenomena. A recent flash sintering review [54] estimates the fractions of FS sintering papers of 50% published on zirconia, 37% on NTC materials and only 10% and 3% on highly insulative and conductive materials, respectively.

In this work, the applicability of the net shape flash spark plasma sintering (NSFSPS) technique to a wide range of materials from metals to electric insulators is presented. Contrary to the previously used sinter-forging flash SPS method [49,51], the NSFSPS approach uses a graphite die electrically insulated by a sprayed boron nitride layer [55] to concentrate the electric current into the sample, if electrically conductive, or in the nearby graphite foil, if electrically insulative. The graphite die allows an easy control of the final shape of the specimen and makes the overall pressure assisted process more stable. The NSFSPS approach uses graphite tools for high temperature applications and good thermal shock resistance. This method is related to the electric current concentration approaches introduced by Zapata-Solvas *et al* [56] for ceramic powders ($ZrB_2$, $MoSi_2$, $Al_2O_3$) in flash conditions and by Román-Manso *et al* [57] for the densification of SiC. However, in contrast with the above-mentioned techniques, the different current patterns employed in this paper demonstrate that an unordinary thermal runaway can be instigated by the electric current assistance for all kinds of the tested materials. The present study includes a numerical simulation to determine the efficiency of the flash heating phenomena for the different materials tested. This study demonstrates the possibility of a flash, homogeneous, energy efficient and near net shape sample sintering applicable to every material regardless of its electric conductivity.

## Results

The current and displacement curves obtained for the Ni, 3Y-$ZrO_2$, and $Al_2O_3$ samples under "constant current mode" and fast electric current ramping conditions (see "forced mode" in section Methods) are reported in figure 1. For each of the displacement curves, the pre-sintering heating generates a thermal dilatation of all the powder volumes responsible for a small negative displacement (indicated by the blue to red arrows). The densification stage (indicated by gray arrows) results in a positive displacement of about 1 mm with duration between 8 and 35 s. In both the "constant current mode" and the "forced mode", the sintering time is smaller than 60 s thereby falling under the flash sintering category [54]. When the displacement plateau appears, the electric current decreases down to zero, and the cooling stage starts. During the cooling stage, a thermal contraction of the whole NSFSPS setup is characterized by a positive displacement (red to blue arrows). The SEM images of the center and the edge and at half the height of all the samples are shown in figure 2. The final relative density and average grain size obtained for all the experiments are given in Table 1. Each sample is nearly dense with a residual porosity between 1 and 3% for the ceramics and 5% for nickel powder. For zirconia and alumina, the grain size increased form an initial 37-100 nm range to about 20 µm in roughly 10 s. This suggests that the fast densification is accompanied by unusually fast grain growth kinetics. The comparison of the transgranular and fractured areas indicates that the residual porosity is present both at the grain boundary and inside the grain volume. The intragranular porosity is particularly pronounced in the metal sample. The comparison of the center/edge SEM images reveals no obvious microstructural differences indicating a very homogeneous process. Nevertheless, a noticeable microstructure difference is observed between the "constant current mode" and the "forced mode" for zirconia. The zirconia sample, that becomes electrically conductive at high temperature, seems to generate some heating instabilities (internal sample temperature runaway and overheating in the sample's area) which cause high grain growth under the "constant current mode" despite both experiments were stopped when the densification reached the plateau. In comparison, the alumina sample, that possesses a low electrical conductivity at low and high temperatures, exhibits nearly no microstructure differences between the two modes. This suggests that the thermal runaway of zirconia enabling the resistive flash sintering process is a source of the unstable grain growth [58] under NSFSPS conditions in contrast with alumina or nickel whose average grain size is nearly unchanged. The non-NTC materials, which are normally impossible to flash sinter using conventional flash methods, appear to be the most stable materials using the NSFSPS method.

The NSFSPS configuration seems to allow a universal fast densification of every material no matter what the material conductivity is. To understand the Joule heating behavior during NSFSPS for each of the different materials employed, a finite element simulation study has been carried out. The materials properties and boundary conditions are described in Ref. [59]. The electric and thermal contact resistances (ECR and TCR) on each interface of the NSFSPS setup are very important parameters influencing the overall temperature distribution [60–64]. Both ECR and TCR are decreased with pressure and temperature and evolve differently depending on the location of the interface and the type of the contact inside the NSFSPS setup [65,66]. The complete determination of each ECR and TCR has been conducted in Ref. [67], those data are used in the present simulations. The temperature, electric current density field, and electric current isolines are reported in figure 3 for each material at the end of the sintering. For nickel powder, the current isolines are concentrated all inside the central column including the sample where the higher temperature is located also in the center. For alumina powder, all the current lines are concentrated in the graphite foil avoiding the electrically insulative sample where the higher temperatures are located at the edge. The electric current distribution is similar for the zirconia powder sample with a small portion of the current that goes through the sample, which becomes conductive at high temperatures but with the values of the electric conductivity still far below the electric conductivity of graphite (103 S/m for zirconia and 1.1E5 S/m for graphite at 1300°C). The zirconia sample starts the self-heating at high temperatures which balances the sample temperature distribution. The TCR, that decreases the heat flowing into the lateral punch/die interface, is the main factor explaining the high temperature concentration inside the NSFSPS setup's central column independently of the sample material nature. The heat generated by the central column is then more easily contained in the sample, which helps the gradients' reduction. The higher the TCR is, the more homogeneous are the sample temperatures and, consequently, the more homogeneous are the respective microstructures. The temperature difference between the sample center and the edge is 140 K for nickel, 93 K for zirconia, and 138 K for alumina, which are the reasonable differences considering the high sample heating rates which can reach 10,000 K/min. It should be noted that the BN coating and bigger punch/die gap can drastically increase the lateral TCR and provide an even higher thermal confinement than the one predicted by the model. As suggested by the homogeneity of the observed microstructures, the real TCR should lead to even more optimistic temperature profiles. The thermal confinement and the short sintering time are favorable conditions to ensure the microstructure homogeneity observed experimentally. The thermal confinement results also from the lateral graphite foil heating. For ceramics sintering cases, a significant amount of the electric current is constrained in the sample/die graphite foil (see figure 3b 3c), therefore the amount of the thermal energy dissipated in graphite foil is high. Under these conditions, the lateral graphite foil can be compared to a susceptor that heats the sample from the edge and considerably stabilizes the sample temperature homogeneity. Concerning the nickel sintering case, the electric current concentration in the lateral graphite foil does not happen because the sample is highly conductive (see the electric current isolines in figure 3a). Thus, the nickel sample does not have this lateral heating and is cooled from the edge leading to a temperature difference of 140 K, which is more intense than the temperature difference in ceramics considering the lower sintering temperatures of metals. The forced thermal runaway generates the sample heating rates of 1700 K/min for nickel and 4300 K/min for alumina. For zirconia, a transition regime occurs at the sample onset temperature of 1250°C where the heating rate passes from 3400 K/min to 12000 K/min. This transition stems from the NTC properties of zirconia generating an intrinsic thermal runaway that is added on top of the imposed electric current amplification. This result explains the difference in the average grain sizes obtained for the two modes in the case of zirconia NSFSPS. This zirconia intrinsic runaway may dramatically influence the grain growth and needs to be controlled by shutting down the process cycle immediately after the full densification of the sample is achieved.

To conclude, the extraordinary possibility to homogeneously flash sinter various kinds of materials, no matter what their electrical conductivity is, with a near net shape capability, has been demonstrated experimentally and theoretically. The applicability of a homogeneous flash heating to any material originates from the two main factors. The first factor is the concentration of the electric current inside the powder sample and its adjacent area. The second factor is the confinement of the heat generated in the sample area by the lateral thermal contact resistance that heightens the sample heating rate and helps homogenize the specimen's temperature spatial distribution. It has been shown that, imposing a variable current pattern, equivalent thermal runaways for metals and insulative samples are easily achievable. The zirconia sample demonstrates a high sensibility to the grain growth at the end of the processing cycle due to the intrinsic thermal runaway phenomenon whose onset temperature is around 1250°C in NSFSPS configuration. Nearly fully dense specimens with homogeneous grain sizes have been obtained using the developed NSFSPS technique.

The new NSFSPS method represents an efficient way for the mass production of small objects with optimal production time and high material performance. The fabrication of large size objects is possible too but will require a specific design of the electric current path in the NSFSPS tooling to balance the amplified thermal gradients. Thus, contrary to the regular flash sintering approaches, the presence of the graphite die opens the prospect of the one-step production of complex shapes similarly to conventional powder metallurgy approaches.

# Methods

**Materials and preparation:** All the experiments were performed using a Spark Plasma Sintering machine (SPSS DR.SINTER Fuji Electronics model 5015). The NSFSPS configuration is reported in figure 4. The main difference of the traditional SPS and NSFSPS configurations is in the inner die interfaces (punch/die and sample/die). In the traditional SPS configuration, a 0.15 mm thick graphite foil is inserted in the vicinity of the punch/die interface (0.2 mm gap). For the NSFSPS configuration two graphite foils are inserted in a 0.4 mm punch/die gap. In order to concentrate the electric current on the sample area, the inner die surface is coated by an electrically insulative boron nitride (BN) spray. The sample can then be hybrid-heated both internally and/or by the nearby graphite foil. Three different powders with different electric conductivities have been selected to test the NSFSPS method. A pure nickel powder (Cerac, Ni 99.9% pure, 5 µm) was selected to represent metals, $Al_2O_3$ (Cerac, $Al_2O_3$ 99.99% pure, 37 nm) to represent electrically insulative materials, and 3Y-$ZrO_2$ (Tosoh- zirconia TZ- 3YS, 100 nm) - as an intermediate electric conductivity material with an NTC behavior. The amount of each used powder introduced has been calculated aiming to obtain 2 mm dense pellets.

**Configuration and current patterns:** The current cycle has been imposed manually in two patterns. A constant pressure of 90 MPa was applied from the start until the end of each experiment. For the ceramics powders, a constant applied current (named "the constant current mode") has been imposed in order to reveal a potential intrinsic thermal runaway during the ceramic sample heating. For the nickel powder, a material without NTC behavior, an electric current cycle with a forced runaway was employed. The electric current was manually increased at 70 A/min up to the beginning of the densification, then the current was quickly increased at about 2500 A/min up to reaching the displacement plateau indicating the end of the densification. We named this electric current pattern "the forced mode" and it was applied also to all the three considered materials. The "forced mode" is a very flexible solution that allows an easy regulation of the electric current value so that the sintering happens under control of the overheating and melting problems with the possibility of stopping the experiment when sintering is finished.

**Main differences compared to the traditional flash sintering:** The first difference is the applied pressure that accelerates and stabilizes the overall sintering process. A die is employed for a better control of the final sample shape (here - cylindrical). For nickel and alumina, the thermal runaway that appears in traditional flash sintering approaches (for NTC materials) is artificially imposed (being not "intrinsic") by an electric current pattern (in the forced mode) that reproduces a similar sample's thermal response corresponding to the material's electric conductivity.

**Acknowledgements**

The support of the US Department of Energy, Materials Sciences Division, under Award No. DE-SC0008581 is gratefully acknowledged.

**Data availability**

The authors declare that the data supporting the findings of this study are available from the corresponding author upon a reasonable request.

**Author contributions**

EO managed the overall project. C.M. and G.L. conducted the SPS experiments. G.L conducted all the SEM characterizations. C.M. performed the finite element simulations. C.M. wrote the paper with input from G.L. and E.O. E.O. refined the paper. All authors contributed to the interpretation of the results.

**Additional information**

Competing interests: The authors declare no competing financial interests.

**Figure captions**

Fig. 1: Experimental electric current (a,c) and displacement (b,d) curves under constant current mode (a,b) and forced mode (c,d).

Fig. 2: SEM images in the centers and edges of nickel, zirconia and alumina samples for constant and forced current modes.

Fig. 3: Simulated temperature, electric current density field, and electric current for nickel, zirconia, and alumina samples.

Fig. 4: Net shape flash spark plasma sintering configuration, the lateral graphite foil is coated with a boron nitride spray to electrically insulate the die and concentrate the electric current on the sample.

**Table captions**

Table 1: Average grain size and relative density and their standard deviation for all the samples.

# Table

Table 1: Average grain size and relative density and their standard deviation for all the samples.

|  |  | Constant current mode | | Forced mode | |
|---|---|---|---|---|---|
|  |  | Edge | Center | Edge | Center |
| Average grain size (μm) | $Al_2O_3$ | 18 ± 5 | 16 ± 4 | 19 ± 7 | 19 ± 5 |
|  | 3Y-$ZrO_2$ | 12 ± 5 | 14 ± 4 | 7 ± 1 | 6 ± 1 |
|  | Ni | - | - | 5 ± 2 | 5 ± 3 |
| Pellet relative density (%) | $Al_2O_3$ | 97 ± 1 | | 97 ± 1 | |
|  | 3Y-$ZrO_2$ | 100 ± 1 | | 99 ± 1 | |
|  | Ni | - | | 95 ± 1 | |

F1

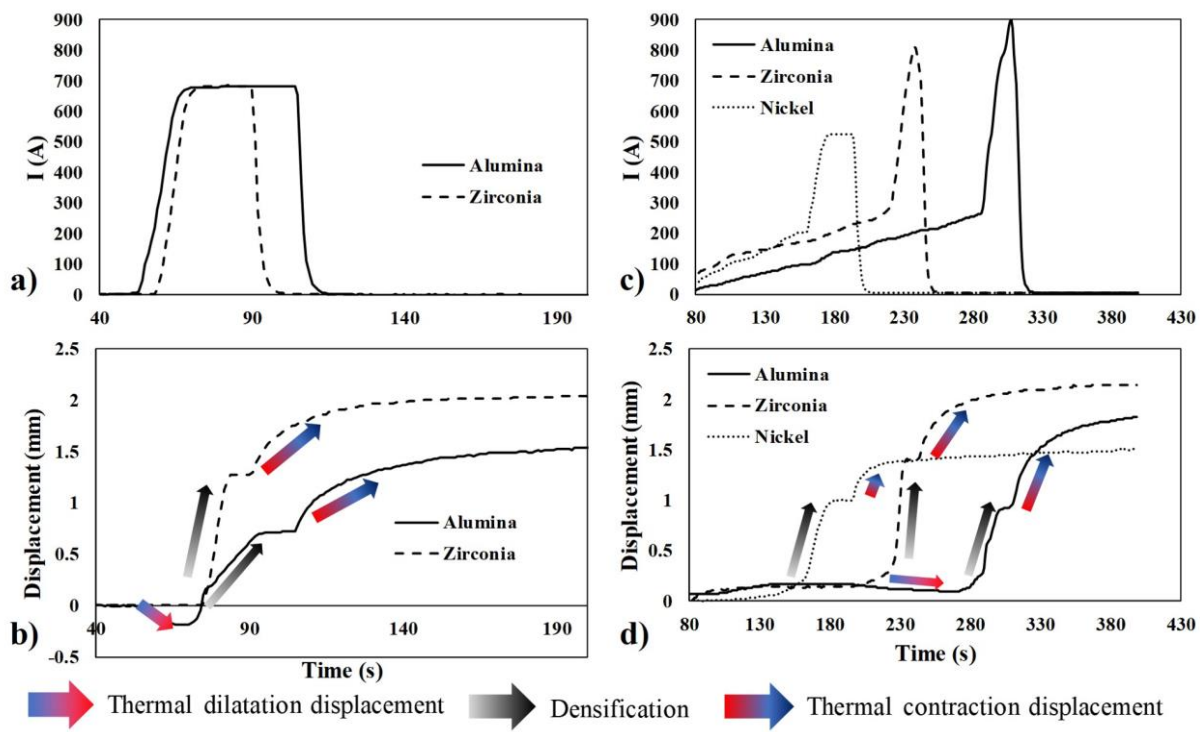

F2

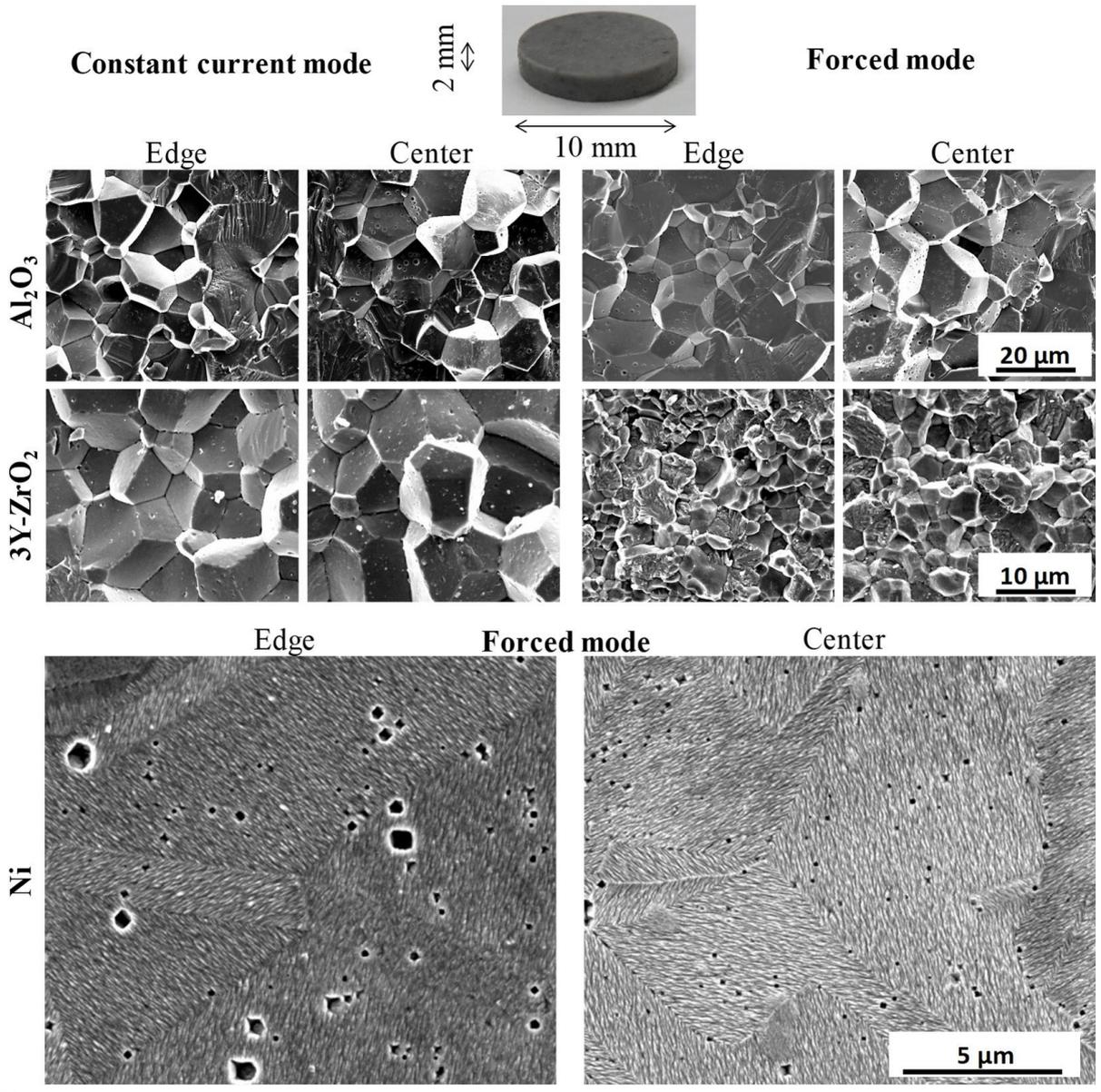

F3

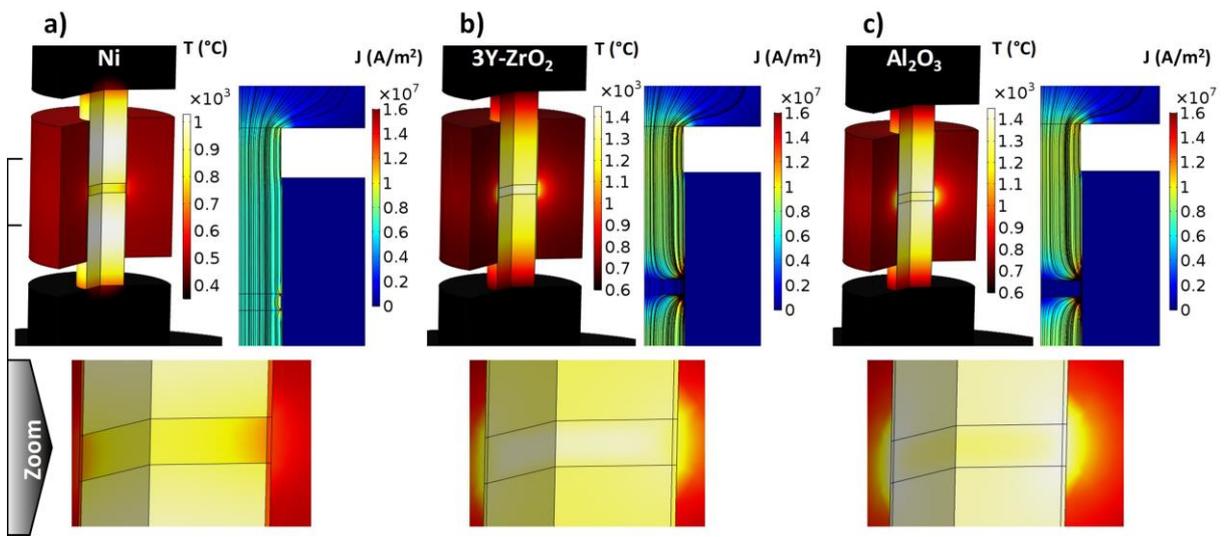

F4

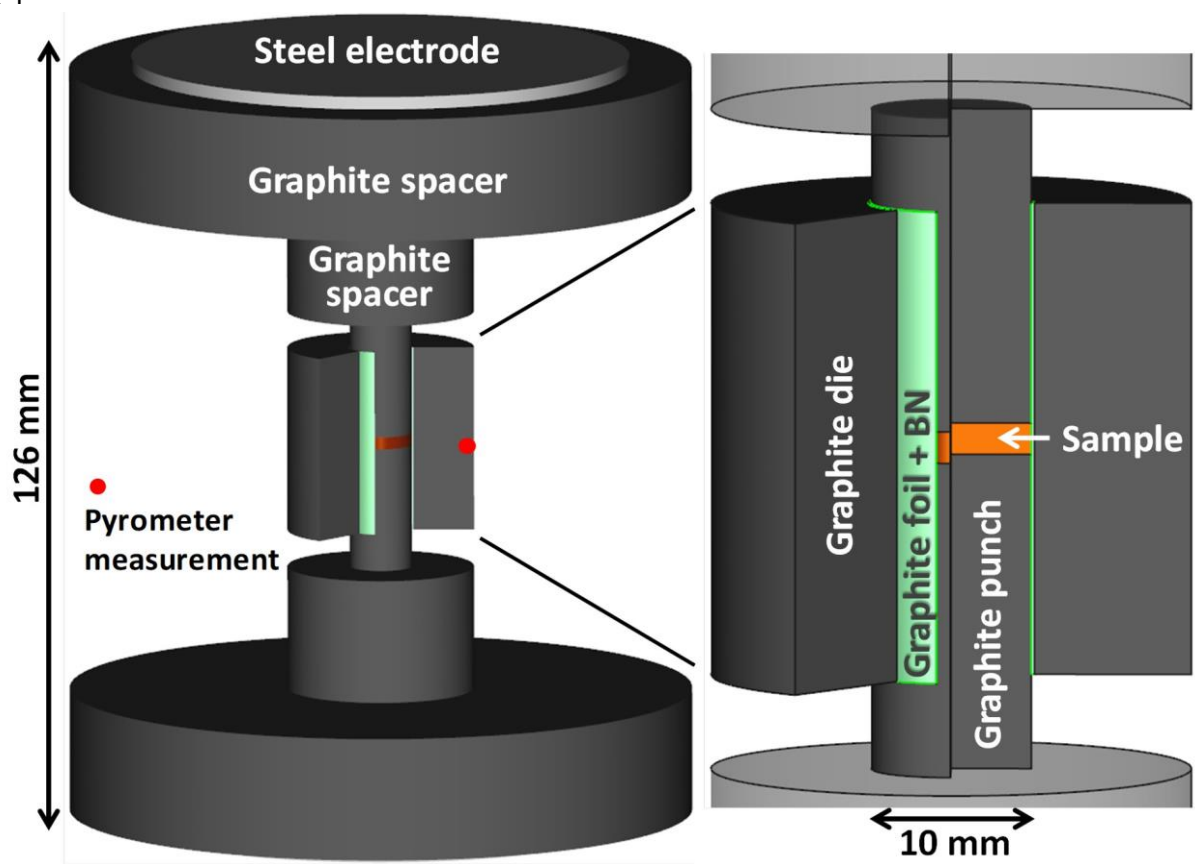